\documentclass[submission,copyright,creativecommons,a4paper]{eptcs}
\providecommand{\event}{FESCA 2014} 
\usepackage{breakurl}             
\usepackage[latin1]{inputenc} 
\usepackage{graphicx}  
\usepackage{amsmath,amsfonts,amssymb,amsthm}
\usepackage{paralist}
\usepackage{stmaryrd}
\usepackage{eucal}
\usepackage{listings}
\usepackage{subfigure}
\usepackage{pgfplots}

\usepackage{psfrag}
\usepackage{epstopdf}

\usepackage{wasysym}

\newcommand{\ms}[1]{\ifmmode%
\mathord{\mathcode`-="702D\it #1\mathcode`\-="2200}\else%
$\mathord{\mathcode`-="702D\it #1\mathcode`\-="2200}$\fi}

\title{Towards Verifying Safety Properties\\of Real-Time Probabilistic Systems}

\author{Fenglin Han
\institute{Norwegian University of Science and Technology, \\Trondheim, Norway}
\email{sih@item.ntnu.no}
  \and
Jan Olaf Blech
\institute{RMIT University, Melbourne, Australia}
\email{janolaf.blech@rmit.edu.au}
  \and
Peter Herrmann
\institute{Norwegian University of Science and Technology, \\ Trondheim, Norway}
\email{herrmann@item.ntnu.no}
  \and
Heinz Schmidt
\institute{RMIT University, Melbourne, Australia}
\email{heinz.schmidt@rmit.edu.au}
      }
\def\titlerunning{Towards Verifying Safety Properties of Real-Time Probabilistic Systems}
\def\authorrunning{F. Han, J.O. Blech, P. Herrmann, H. Schmidt}

\begin{document}

\maketitle

\begin{abstract}                          %
Using probabilities in the formal-methods-based development of safety-critical software has quickened interests in academia and industry.
We address this area by our model-driven engineering method for reactive systems SPACE and its tool-set Reactive Blocks that provide an extension to support the modeling and verification of real-time behaviors.
The approach facilitates the composition of system models from reusable building blocks as well as the verification of functional and real-time properties and the automatic generation of Java code.

In this paper, we describe the extension of the tool-set to enable the modeling and verification of probabilistic real-time system behavior with the focus on spatial properties that ensure system safety.
In particular, we incorporate descriptions of probabilistic behavior into our
Reactive Blocks models and integrate the model checker PRISM which
allows to verify that a real-time
system satisfies certain safety properties with a given probability.
Moreover, we consider the spatial implication of probabilistic system specifications by integrating the spatial verification tool BeSpaceD and give an automatic approach to translate system specifications to the input languages of PRISM and BeSpaceD.
The approach is highlighted by an example.

\end{abstract}

\section{Introduction}

Modeling and verification methods for embedded control system in
domains such as avionics, automotive and robotics should address a
variety of software and hardware aspects including real-time and
probabilistic properties, distribution of system components, communication protocols, characteristics of digital circuits and controllers.
Real-time systems can require quantitative timing constraints which may include guaranteed probabilities for time and spatial properties. For example, a robot may be required to process a task in a predefined amount of time with a probability of 99.999999\% to prevent expensive maintenance operations resulting from minor damage to the equipment. It must complete the task in a slightly larger amount of time with 100\% to prevent major damage.

Here, we propose a framework for integrating probabilistic real-time verification and performance prediction with system development.
This approach extends our existing model-driven engineering framework SPACE and its tool-set Reactive Blocks\footnote{Until recently, Reactive Blocks was called \emph{Arctis}.}~\cite{KrSH:JSS09} with real-time system behavior verification and schedulability analysis \cite{HaHL:ICECCS13,HaHe:UMedia13}.
Reactive Blocks enables the model-based engineering of reactive systems by composing reusable building blocks each describing a certain sub-functionality of a system.
The composed system model is automatically transformed into executable Java code~\cite{KrSH:JSS09}.
Further, various formal verification methods ensure functional correctness~\cite{KrSH:JSS09} as well as reliability~\cite{SlKH:GPCE11}, security~\cite{GuHe:ESSOS2013} and safety~\cite{HaHL:ICECCS13,HaHe:UMedia13} of targeted systems.

The formalism for modeling  probabilistic real-time systems used in
this work is based on probabilistic timed automata (PTA) \cite{KNSW04}.
It incorporates both probabilistic and real-time
characteristics. Probabilistic properties are represented with an extension of Computational Tree Logic, i.e., PCTL~\cite{HJ94}.
PRISM \cite{{KNP11}} is a probabilistic model checker for formal analysis of systems
that exhibits stochastic behaviors. It supports multiple-formalisms,
including discrete-time Markov chains, continuous-time Markov chains,
Markov decision processes and PTA
making it possible to capture the random behavior of our real-time
system model \cite{HaHL:ICECCS13}, for example random aspects of failure or uncertain inputs, loads or timing. We choose PTA for the stochastic
behavior  since it integrates well with our existing timed-automata based real-time system model.
PTA has an equivalent descriptive power to a MDP (Markov Decision Process) \cite{Der70} such that we can use the terminology of MDP for the system descriptions.
Probabilistic CTL \cite{KNP07a} has the capability of expressing real-time as well as probabilistic properties of a reactive system.

The tool set described in this paper involves the five engineering steps outlined below:
\begin{enumerate}
\item
We model a system using Reactive Blocks including a simulator of its environment, in particular the spatial conditions to be reflected. In this model, we annotate
probabilities as well as real-time behavior.
\item The model is analyzed with the model checker built into Reactive Blocks for functional errors and transformed into an executable simulator.
\item
The simulator is carried out and, during the simulation, traces capturing spatiotemporal behavior
with or without annotated
probabilities are extracted for spatial verification.
\item The extracted spatiotemporal behavior is verified for possible
  spatial implications like collisions using our BeSpaceD tool \cite{issec}.
  Here, distributions capturing the combined probabilistic time behavior of the subcomponents are created from the
  extended Reactive Block models using the PRISM-based analysis.
\item If all analyses are passed, the simulator sub-functionality is removed from the Reactive Blocks model such that its core functionality can be transformed into executable code.
\end{enumerate}
In this paper, we have three main contributions.
\begin{enumerate}
\item A novel approach for system performance predictions is introduced.
In particular, we present a probabilistic real-time state-machine
for software component performance descriptions.
This so-called PRTESM is an extension of the External State Machines (ESMs)~\cite{KrHe:MODELS09} used in Reactive Blocks.
It allows to express probabilistic real-time assumptions and guarantees of a building block.
That enables us to compose the PRTESMs of the various building blocks forming a system to predict probabilities of the overall system behavior.
\item We show the integration of the model-checker PRISM~\cite{KNP11} to the tool-set.
For that, the PRTESMs are transformed into PTAs~\cite{KNSW04} and the performance predictions of the overall system to PCTL statements~\cite{HJ94} which can be directly proven by PRISM.
\item We look at spatial implications of probabilities in system behavior.
The introduction of known probabilities into system behavior allows us
to calculate how likely a physical unit like a robot will be present
in a given area in space. We use the tool BeSpaceD \cite{issec} for
this.
\end{enumerate}

\subsection{Guiding Example} \label{Sect:GuidingExample}
\begin{figure}[tbp]
\begin{center}
\includegraphics[width=0.7\textwidth]{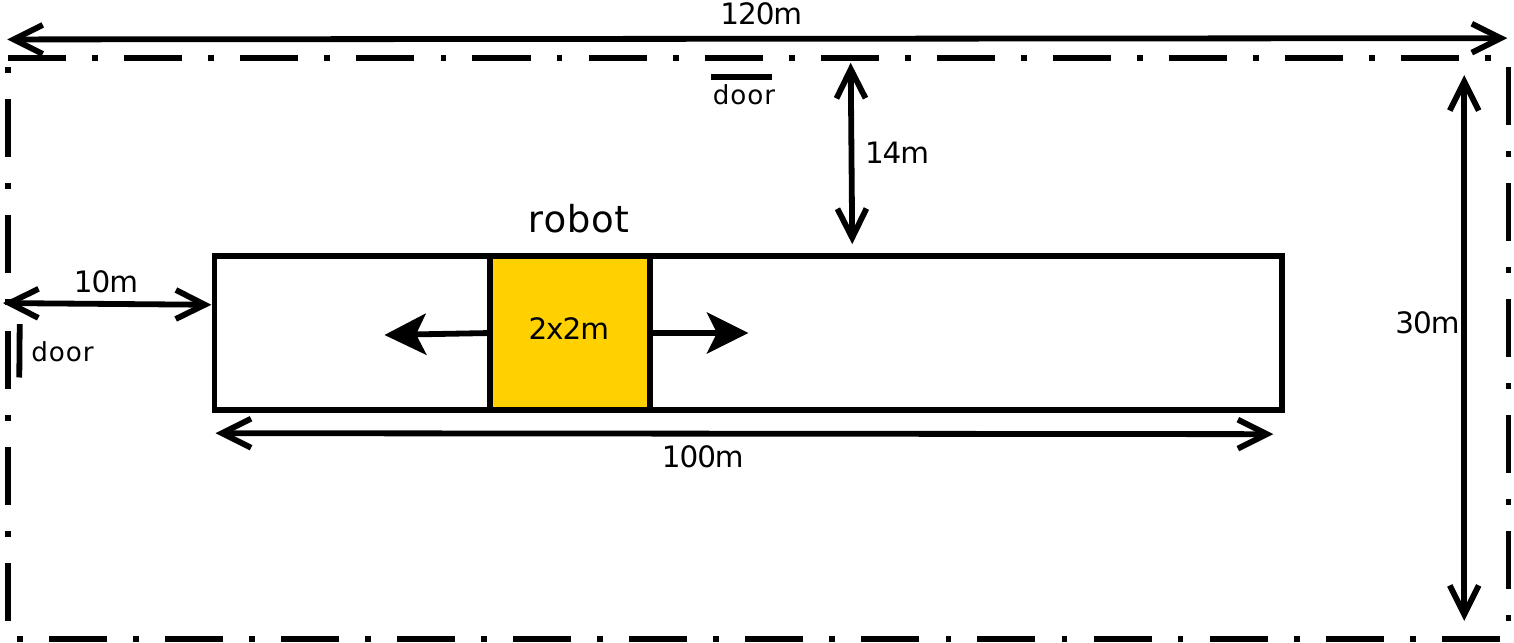}\hspace{5mm}
\caption{Layout of the moving robot}
\label{fig:movingRobot}
\end{center}
\end{figure}
We illustrate the tool set by a scenario of a moving robot in a
factory hall featuring probabilistic behavior. Figure
\ref{fig:movingRobot} provides a spatial layout of the example
scenario. The moving robot occupies a 2 x 2
meters space in the 120 x 30 meters factory hall and moves along a
straight line in the center of the room covering a distance of 100
meters. The maximum speed of the robot can reach 10 $\frac{m}{s}$ and
thus a collision with a human may lead to fatal injuries. To eliminate such injuries,
the hall is equipped with sensors monitoring the
robot for approximations of humans.
If the robot comes close to a human, it is slowed down or even stopped.
The probabilistic aspect of this example comprises probability distributions on the reaction time once
a human is detected.
In this paper, the robot controller and a simulator of the continuous robot behavior are developed and implemented using Reactive Blocks.

\subsection{Overview}

The paper is arranged as follows: In Section~\ref{label:sb}, we give a
 description of the model of the robot control system example in
 Reactive Blocks realizing the distributed control functions as well
 as the robot simulation. %
Section~\ref{sec:prtesm}
 introduces our formalism for probabilistic time constraint and the
 translation into PRISM input. Section~\ref{sec:spatial} presents our approach for spatial implications of probabilistic system behaviors and introduces the tool for probabilistic spatiotemporal property verification.
We present the verification of our  properties in Section~\ref{sec:result}.
Related work is discussed in Section~\ref{label:relatedwork} followed by a conclusion in Section~\ref{label:conclusions}.

\section{Modeling Control Functions}\label{label:sb}
In the moving robot example sketched in
Section~\ref{Sect:GuidingExample}, the three main activities are:
\begin{enumerate}
\item
Polling of sensor data about the positions of the robot and the
human\footnote{For simplicity, we only consider the human closest to the robot.}.
\item
Deciding the correct operation mode of the robot based on the
distance to the human.
\item Forwarding an altered operation mode to the robot controller.
\end{enumerate}
All three activities together should be performed within $0.5~s$ at maximum.

We model these control functions as well as the simulator of the human and robot behaviors separately from each other by different building blocks.
In SPACE and Reactive Blocks, a building block consists of a behavioral model in the form of a UML activity~\cite{KrSH:JSS09} supplemented by an External State Machine (ESM)~\cite{KrHe:MODELS09} describing its interface behavior.

\begin{figure}[tb]
\begin{center}
\includegraphics[width=\textwidth]{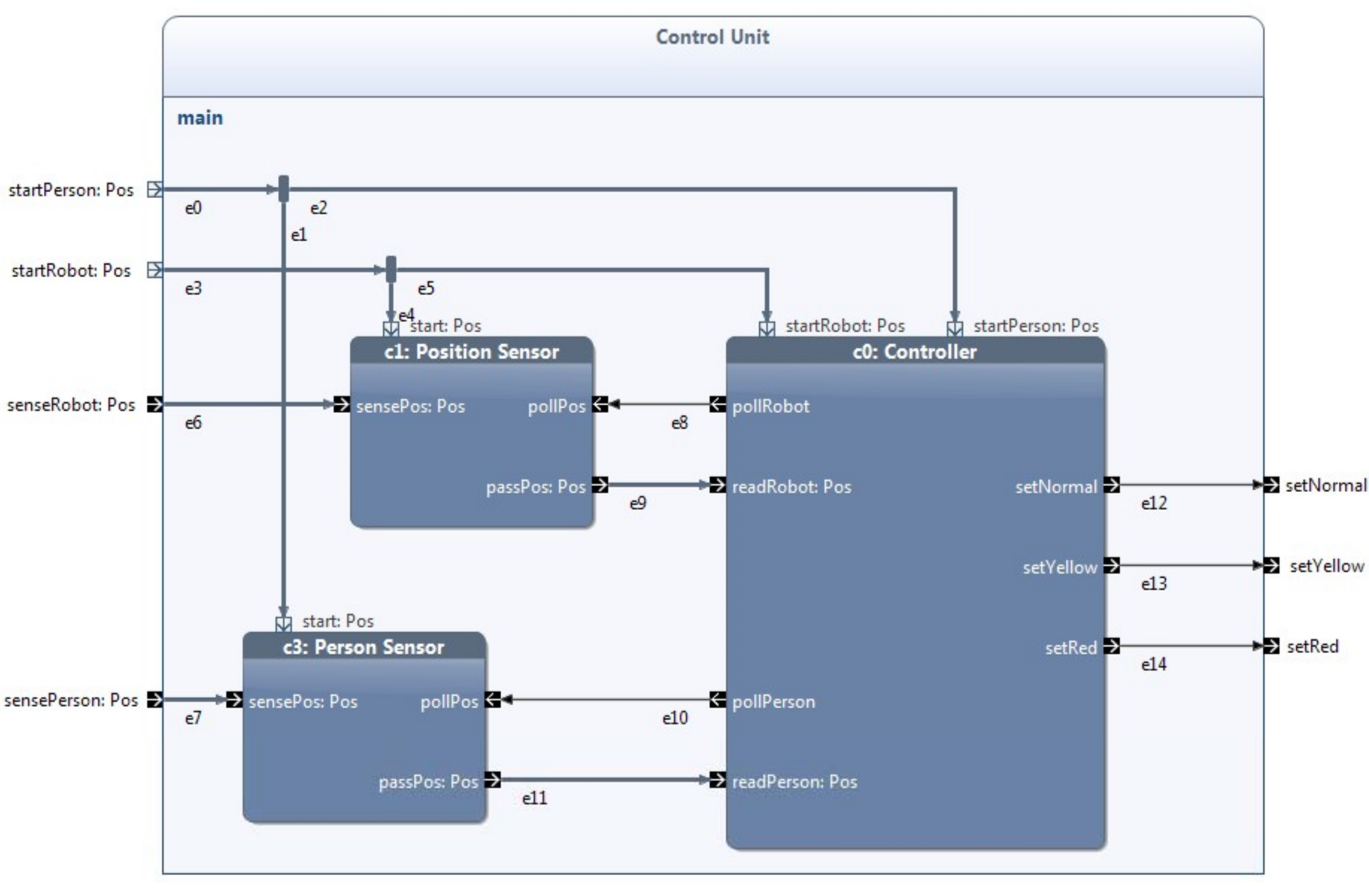}
\end{center}
\caption{UML activity of building block \emph{Control Unit}}
\label{fig:controlunit}
\end{figure}
The safety controller protecting humans from collisions with the fast moving robot is specified by the building block \emph{Control Unit}.
The UML activity modeling the behavior of this block is shown in Figure~\ref{fig:controlunit}.
Similar to Petri Nets, the control and data flows are represented by the flow of tokens in the activities. These tokens are passed by the activity edges towards vertices. Vertices can be control elements (such as forks duplicating tokens) or operations (associated with Java methods executed when a token arrives).
Further, activities may contain call behavior actions like \emph{Controller}, \emph{Position Sensor} and \emph{Person Sensor} each referring to another building block.
The interaction between the activity containing a call behavior action of a building block and the one referring to its behavior is modeled by pins and parameter nodes.
Parameter nodes are the identifiers at the edge of an activity, e.g., \emph{startRobot} in block \emph{Control Unit}.
All parameter nodes of an activity are available as pins in the call
behavior actions referring to its building block (e.g., \emph{setRed}
in block \emph{Controller}).
The semantics defines that a token reaching a pin of a call behavior action continues from the corresponding parameter node of the activity referring to the behavior of the call behavior action and vice versa.

\emph{Position Sensor} and \emph{Person Sensor} refer to the sensors for the positions of the robot and human which get the current position information from the simulation via their input pins \emph{sensePos}.
The block \emph{Controller} realizes the safety controller of the system.
It polls the robot and human positions every $10~ms$ using the pins \emph{poll} and \emph{read}.
From these inputs, the distance between human and robot is computed and the correct operation mode is selected.
Altogether, there are three operation modes that are defined as follows:
\begin{itemize}
   \item \emph{Normal mode:} If no human is closer than 25 meters to the robot, the robot accelerates with $5\frac{m}{s^2}$ until reaching a speed of $10\frac{m}{s}$.  When it is $11~m$ close to its endpoint, it decelerates with $5\frac{m}{s^2}$ until reaching a speed of $1\frac{m}{s}$ which is carried until reaching the endpoint.
   \item \emph{Yellow mode:} If a human is detected in a distance of less than 25 meters but more than 10 meters, the robot is slowed down with a rate of $10\frac{m}{s^2}$ until reaching a speed of $2\frac{m}{s}$ (resp. $1\frac{m}{s}$ if it is closer than $11~m$ to its endpoint).
   \item \emph{Red mode:} If the human is within 10 meters range to the robot, the robot makes an emergency stop with a deceleration of $15\frac{m}{s^2}$.
\end{itemize}

\begin{figure}[tb]
\begin{center}
\subfigure[ESM of \emph{Control Unit}. \label{fig:controlUnitESM}]{
\includegraphics[width=0.3\textwidth]{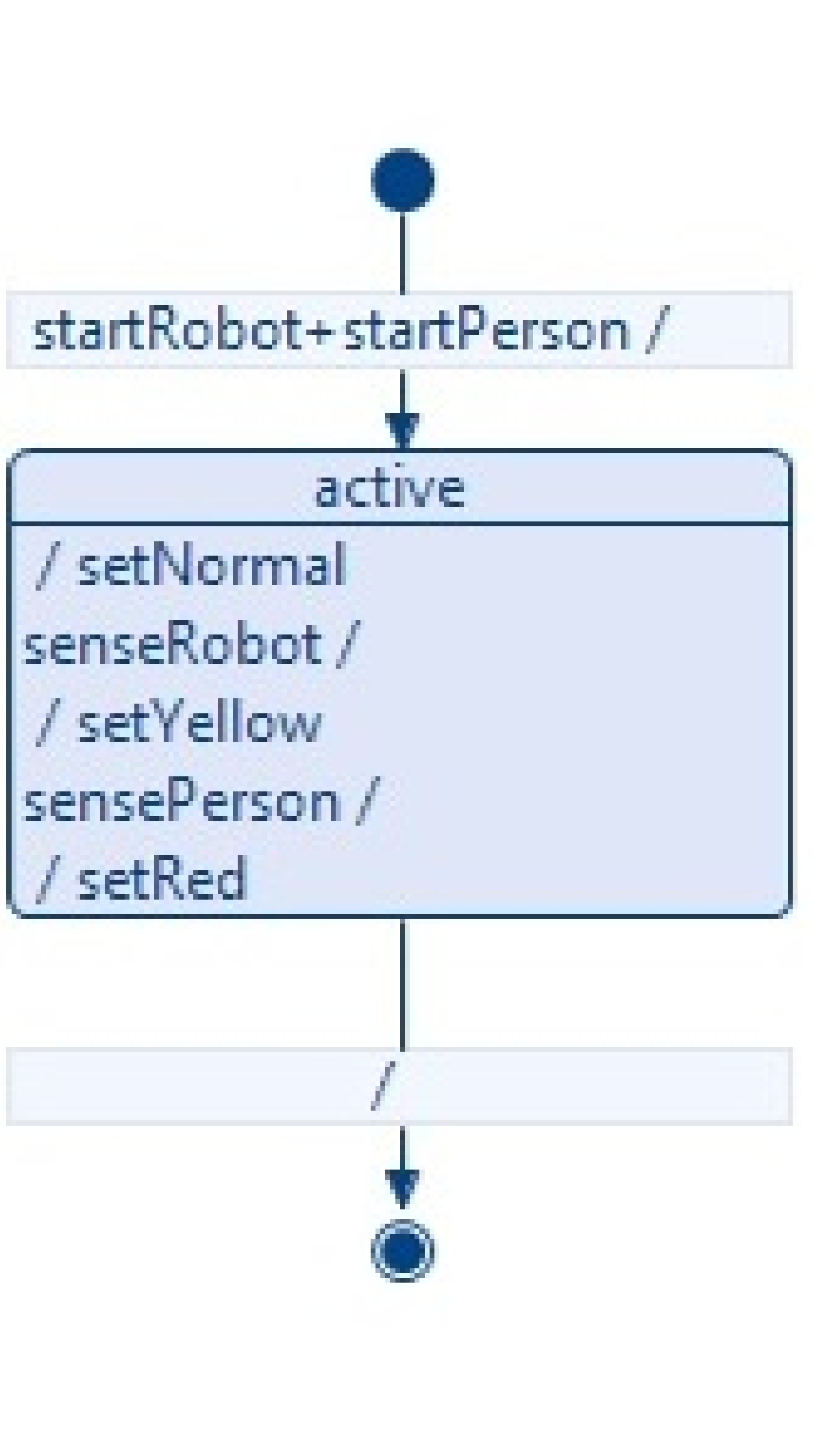}}\hspace{5mm}
\subfigure[ESM of \emph{Robot Operation}. \label{fig:robotOperationESM}]{
\includegraphics[width=0.3\textwidth]{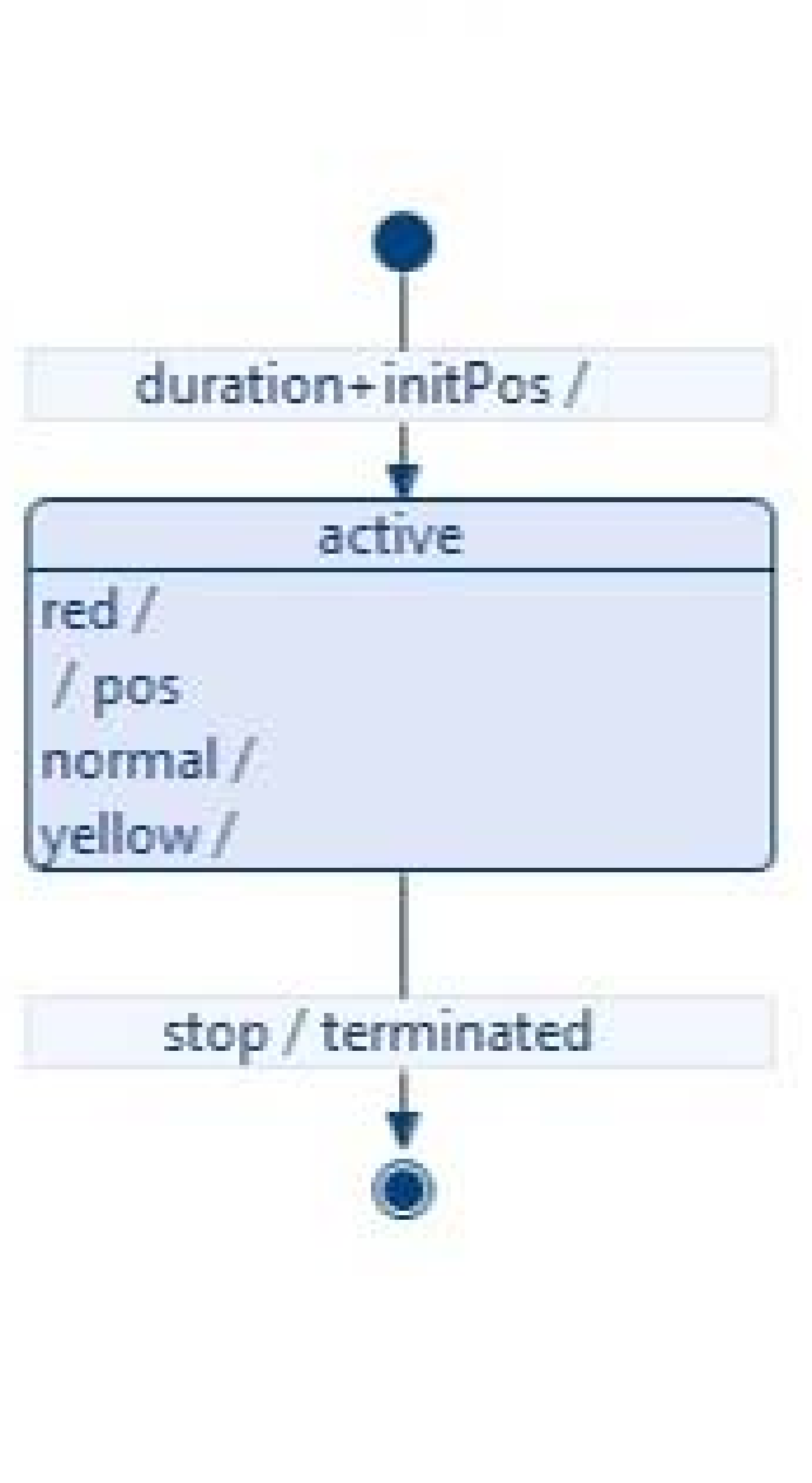}}
\end{center}
\caption{ESMs of building blocks \emph{Control Unit} and \emph{Robot Operation}}
\label{fig:ESMs}
\end{figure}
The behavior of the building block \emph{Control Unit} is specified by its ESM depicted in Fig.~\ref{fig:controlUnitESM}.
An ESM is a UML state machine describing which of its parameter nodes are passed by tokens in a certain transition.
\emph{Control Unit} is initiated by parallel flows through the parameter nodes \emph{startRobot} and \emph{startPerson} which contain the initial positions of robot and human.
Thereafter, the building block is in state \emph{active} in which the environment (symbol $/$ right of the parameter node identifier) may send position data via parameter nodes \emph{senseRobot} and \emph{sensePerson} while from the block itself ($/$ left of the parameter node identifier) the current operation mode may be sent via \emph{setNormal}, \emph{setYellow} and \emph{setRed}.
The building block is terminated and all remaining tokens on its activity are removed if the activity containing the call behavior action of the block is terminated as well which is described by the transition $/$.

\begin{figure}[tb]
\begin{center}
\includegraphics[width=\textwidth]{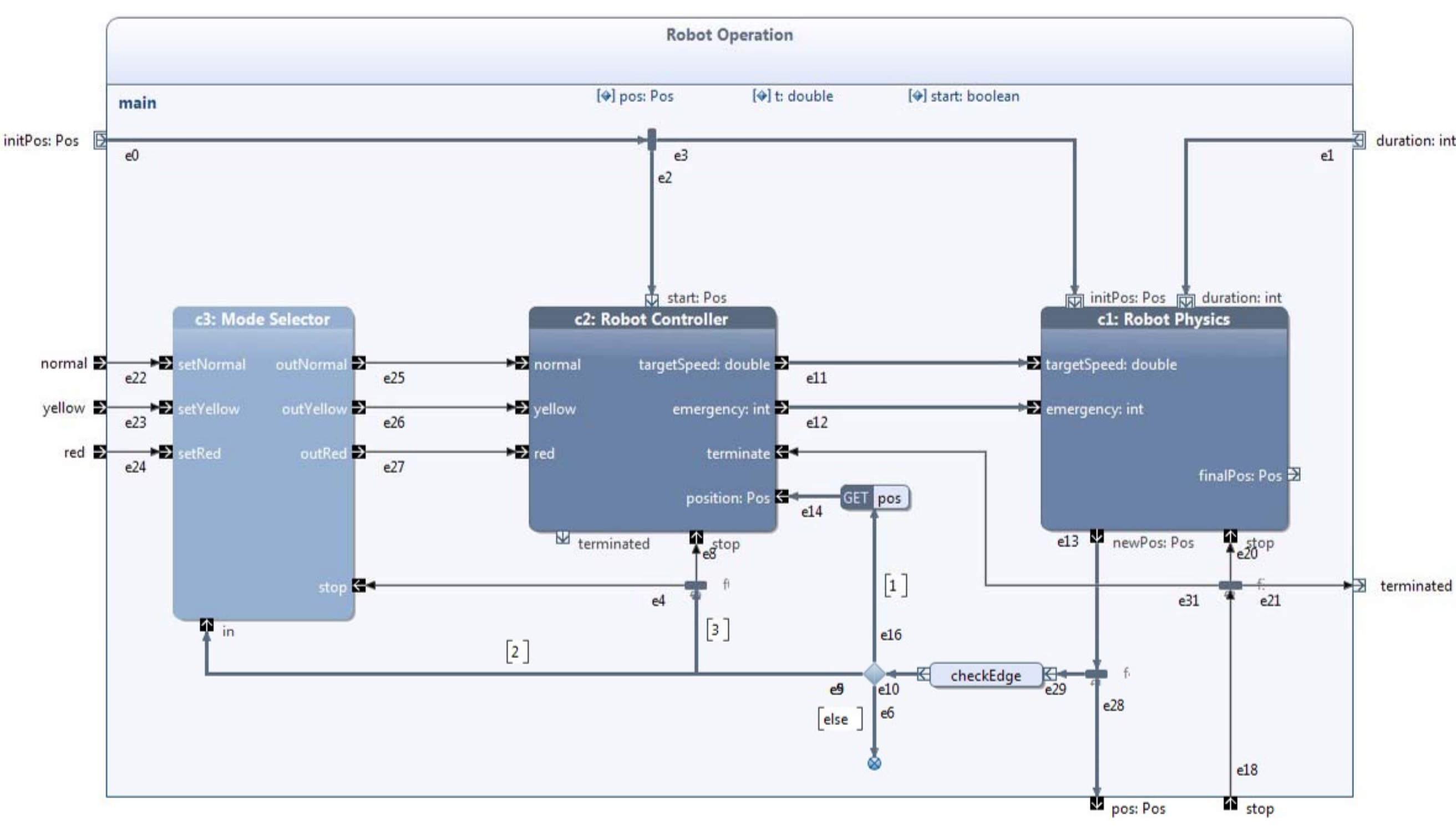}
\end{center}
\caption{UML activity of building block \emph{Robot Operation}}
\label{fig:robotoperation}
\end{figure}
Figure~\ref{fig:robotoperation} shows the UML activity of building block \emph{Robot Operation}.
It contains the building block \emph{Mode Selector} storing the current operation mode.
\emph{Robot Controller} models the controller of the robot, which chooses the current robot speed according to the operation mode and the position of the robot in the factory hall.
Using the block \emph{Robot Physics} we model the simulator for the robot.
The continuous behavior is specified by a difference equation which is executed every $5~ms$.

The ESM of block \emph{Robot Operation} in Figure~\ref{fig:robotOperationESM} determines that the block is started by parallel token flows via the parameter nodes \emph{duration} and \emph{initPos} which refer to the execution time of the difference equation (i.e., $5~ms$) and the initial position of the robot.
In state \emph{active}, the operation modes \emph{normal}, \emph{yellow} or \emph{red} may be received by the environment while by \emph{pos} the current position of the robot may be forwarded towards the sensor in block \emph{Control Unit}.
The block is terminated by a token coming via parameter node \emph{stop} which leads to an output via \emph{stopped}.
This signal leads to the termination of all inner blocks followed by the termination of \emph{Robot Operation}.

The third block of the example is called \emph{Suicidal Human}.
This somehow odd name refers to the simulation of a human attempting to approach the robot as fast as humanly possible.
So, it describes the worst case situation to be solved by the safety controller.
Like the robot, the behavior of the human is specified by a difference equation that is executed every $5~ms$.
For the sake of brevity, we do neither list this block nor the other blocks of our system in detail here.

After creating and composing all building blocks of our system, we can check them for the presence of functional design errors like not fulfilling their ESMs (see~\cite{KrSH:JSS09}).
Further, if all checks are passed, Reactive Blocks automatically generates executable Java of our system which can be carried out to simulate robot runs.

\subsection{Probability Assumptions and Temporal Safety} \label{label:assumption}
As discussed above, the factory hall, in which humans and machines collaborate in close proximity, is
monitored by camera sensors for collision avoidence.
A safety controller constantly monitors the operation for the proximity of humans and then decides which operation mode to choose.
Of course, to avoid collisions when the robot is still moving, we have to guarantee maximum reaction times for the different functions carried out in order to slow down or stop the robot.
The four main sub-tasks are the fetching of sensor data including the delay between two pollings of the sensors, the processing time of the safety controller in order to compute the distances between human and robot and to decide about the correct operation mode, the communication delay between the safety and the robot controller as well as the processing time of the robot controller including delays within the robot starting to break.

\begin{table}[tb]
\caption{Accumulative probability distribution of the execution times for the different tasks}
\centering
\begin{tabular}{|l|r|r|r|r|}
\hline
{\bf Delay Type } & {\bf Maximum Time} & {\bf Probability} & {\bf Fig.~\ref{fig:controUnit_prtesm}} & {\bf Fig.~\ref{fig:RobotOperation_prtesm}}\\
\hline \hline
Time to fetch sensor data including & $15~ms$ & $10~\%$ & & \\
polling delay & $17~ms$ & $40~\%$ & &\\
& $18~ms$ & $85~\%$ & &\\
& $19~ms$ & $99.998~\%$ & &\\
& $20~ms$ & $100~\%$ & &\\
\hline
Processing time recognition unit &  $250~ms$ & $10~\%$ & & \\
& $260~ms$ & $30~\%$ & & \\
& $270~ms$ & $60~\%$ & & \\
& $280~ms$ & $90~\%$ & & \\
& $285~ms$ & $99~\%$ & & \\
& $290~ms$ & $100~\%$ & & \\
\hline
Communication time recognition &  $15~ms$ & $80~\%$ & r1 &\\
unit to robot & $16~ms$ & $98~\%$ & r2 &\\
& $16.5~ms$ & $99.5~\%$ & r3 &\\
& $16.9~ms$ & $99.99999995~\%$ & r4 &\\
& $20~ms$ & $100~\%$ & r5 &\\
\hline
Internal robot processing time and & $150~ms$ & $5~\%$ & & r1\\
actuator reaction & $159~ms$ & $90~\%$ & & r2\\
& $160~ms$ & $95~\%$ & & r3\\
& $165~ms$ & $99.9995~\%$ & & r4\\
& $170~ms$ & $100~\%$ & & r5\\
\hline
\end{tabular}
\label{table:t1}
\end{table}
For elaborating our approach, we assume that the probabilities for the
reaction times associated with the sub-tasks correspond to the
percentages depicted in Table~\ref{table:t1}.
They show the probability that a task is finished within a certain time in an accumulative way.
For instance, according to the table, the fetching of the sensor data is finished within $15~ms$ with a probability of $10~\%$ while the overall probability that it is completed within $17~ms$ is $40~\%$.
Thus, we guarantee that this task is carried out within $20~ms$.
Of course, in practice one cannot give axiomatic guarantees of real-time properties since a control system is always subject to external influences like a failure of the computer hardware running it.
We decided to ignore these kinds of external error in our models but
are aware that, when our tool chain is used for real hazard analysis,
such faults have to be taken into account as well. The values in the
table do not correspond to an existing system, but rather represent
typical values one might expect in some field-bus based systems.

\section{Probabilistic Real-Time Extended State Machines}\label{sec:prtesm}

Following the concept of Timed Automata~\cite{AlDi:90}, we
extended our external state machines (ESM) to Real-Time ESMs
(RTESM) in~\cite{HaHL:ICECCS13,HaHe:UMedia13}.
RTESMs allow the specification of deadlines for the time a building block may rest in a certain RTESM state.
As a new contribution in this paper, we introduce the further extension of the RTESMs to Probabilistic
Real-Time External State Machines (PRTESM).
Extending the habitual pattern in Reactive Blocks to model functional and non-functional interface properties of building blocks, the PRTESMs make the
description of probabilistic real-time behavior possible and allow to describe discrete probability distributions
like the ones listed in Table~\ref{table:t1}.
PRTESMs allow a straightforward transformation into Probabilistic Timed
Automata (PTA)~\cite{KNSW04} that can be model checked by verification tools like PRISM \cite{KNP11,KNSS02}.
\begin{figure}[tb]
\centering
\includegraphics[width=0.9\textwidth]{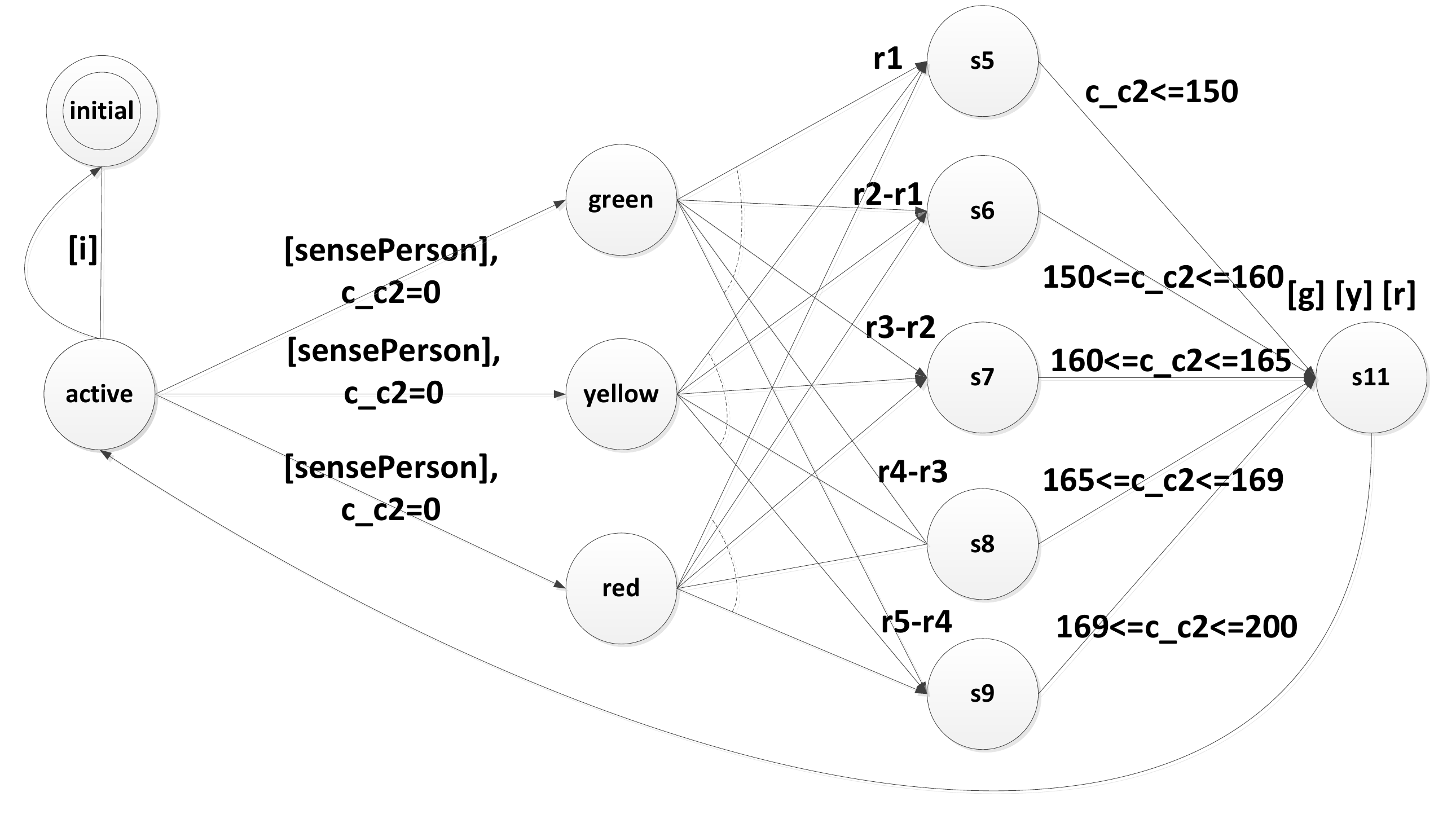}
\caption{PRTESM for the $Control Unit$ block}
\label{fig:controUnit_prtesm}
\end{figure}

\begin{figure}[tb]
\centering
\includegraphics[width=0.9\textwidth]{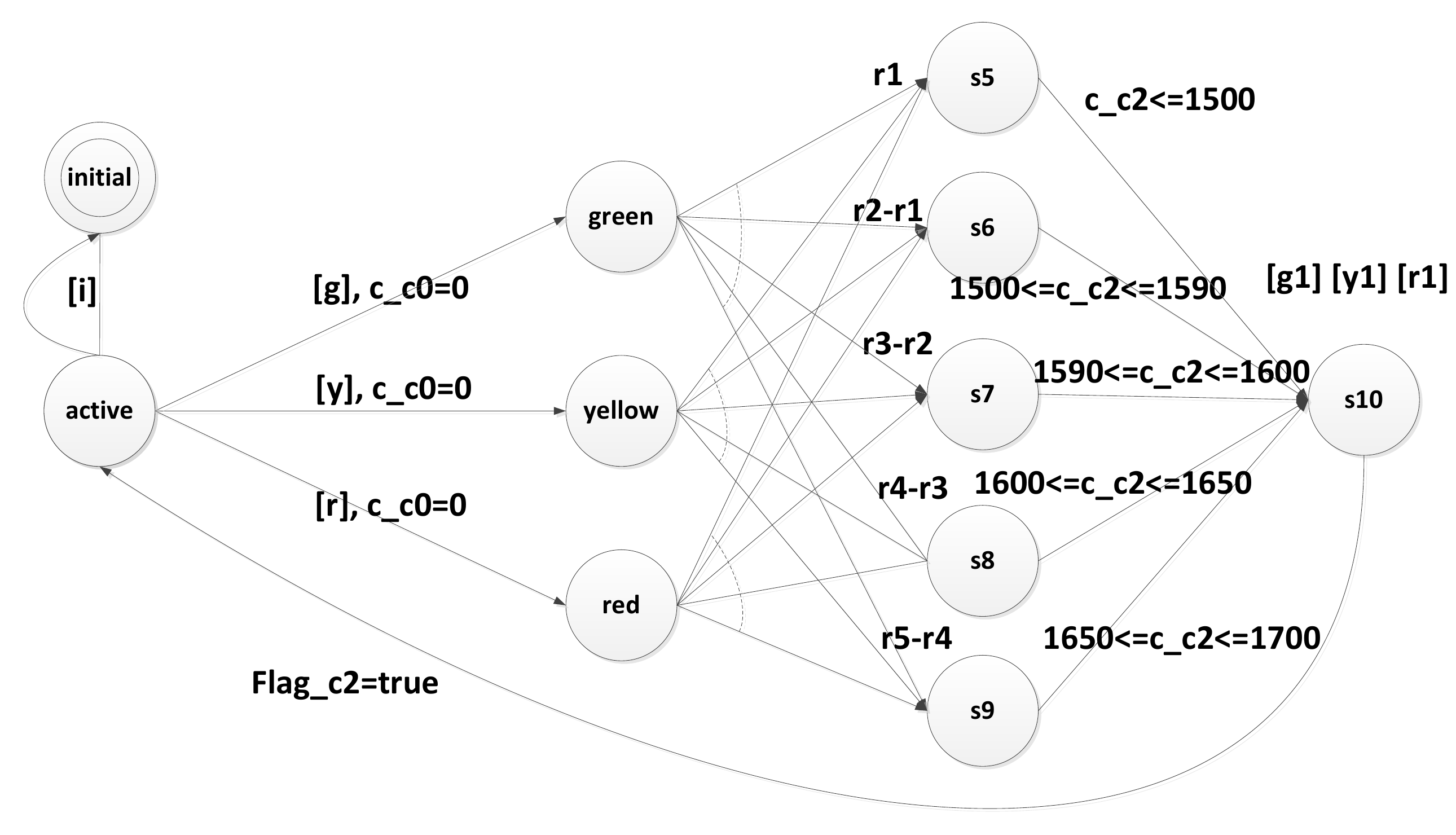}
\caption{PRTESM for $Robot Operation$ block}
\label{fig:RobotOperation_prtesm}
\end{figure}

Figures~\ref{fig:controUnit_prtesm} and \ref{fig:RobotOperation_prtesm} show the PRTESMs of the blocks \emph{Control Unit} and \emph{Robot Operation}.
To facilitate the transformation into PTAs, a PRTESM contains an initial state \emph{initial} representing both the initial and final states of the corresponding ESM.
Moreover, the PRTESM may contain special states that express
probabilistic behavior of the concerned actions as well as the
synchronization semaphores and timed constraints used to model
real-time properties. In the PRTESMs listed in Figures~\ref{fig:controUnit_prtesm} and \ref{fig:RobotOperation_prtesm}, the values {\bf r1}, {\bf r2}, {\bf r3}, {\bf r4},
{\bf r5} represent the probabilities from the third resp. forth section in Table~\ref{table:t1}.
The time deadlines are measured in 100 microseconds.

The approach for the generation of a PTA from a PRTESM for a real-time blocks is
semi-automatic.
\begin{itemize}
\item
First, a set of communication actions are identified in the building
block concerning the underneath hardware or communication protocol.
In our example in the building block \emph{Control Unit}, the
parameter pins \emph{setNormal, setYellow, setRed} and
\emph{sensePerson} realize the communication among distributed agents
in the moving robot scenario. The \emph{setX} set of parameters
realize communication between robot controller and robot actuator
while the \emph{sensePerson} parameter realizes the communication
between camera sensor and robot controller. Thus we extract the
probabilistic real-time actions in the system to PRTESM and ignore
other actions.
\item
Second, transitions corresponding to distributed communications are
transformed to new states and transitions
expressing probabilities. In our example these are: ${active
  \rightarrow_{setNormal} active}$, \\ ${active \rightarrow_{setYellow} active}$, ${active \rightarrow_{setRed} active}$ and ${active \rightarrow_{sensePerson} active}$.
\end{itemize}

\begin{figure}[tb]
\begin{lstlisting}[basicstyle={\small\ttfamily},frame=lines]
1. pta
2.  const int c2_1 = 150; // time unit 0.0001 s
3.  const int c2_2 = 160;
4.  ...
5. const double r1 = 0.8;  // probability
6. const double r2 = 0.98; // accumulative probability
7. module c2_Control_Unit_prtesm
8.         s_c2 : [0..10] init 0;
9.         c_c2 : clock;
10.        flag_c2 : bool init false;
11.        [i] s_c2=0 -> (s_c2'=1);
12.        [r] s_c2=1 -> (s_c2'=2)&(c_c2'=0);
13.        ...
14.        [sensePerson] s_c2=1 -> (s_c2'=2)&(c_c2'=0);
15.        [] s_c2=2 -> r1 : (s_c2'=5) + r2-r1 : (s_c2'=6)
16.        + r3-r2 : (s_c2'=7) + r4-r3 : (s_c2'=8)
17.        + r5-r4 : (s_c2'=9);
18.        [y] s_c2=5&c_c2<=c2_1 -> (s_c2'=10);
19.        [y] s_c2=6&c_c2>=c2_1&c_c2<=c2_2 -> (s_c2'=10);
19.        ...
20.        [] s_c2=10 -> (s_c2'=10) & (flag_c2'=true);
21.endmodule
\end{lstlisting}
\caption{Excerpt of PTA codes corresponds to Figure~\ref{fig:controUnit_prtesm}. \label{fig:code}}
\end{figure}
The code excerpt in Figure ~\ref{fig:code} corresponds to the
PRTESM in Figure~\ref{fig:controUnit_prtesm} and
illustrates the corresponding Probabilistic Timed Automata (PTA) of the building block \emph{Control Unit}. The formalism declaration \emph{pta} demands that the following modules follow the timed automata style.
Time and probability values are declared as constant values before the module declaration. \emph{c2\_Control\_Unit\_prtesm} is the module name.
A PTA transition in PRISM is started with a pair of brackets
(\emph{[]}) and optional synchronization commands in the brackets,
e.g., a semaphore \emph{i} initializes distributed building blocks
simultaneously during the system initialization, and semaphores \emph{r,y}
are abbreviated from communication parameter \emph{setRed, setYellow}
to synchronize module \emph{Control Unit} (robot controller) and
module \emph{Robot Operation} (robot actuator). ESM transitions which
are labeled as real-time probabilistic actions are exported and
extended with probabilistic description. Line \emph{14} to \emph{17}
gives an example command in PRISM showing the probabilities. It
declares that when \emph{s\_c2} variable equals to 2 it has
\emph{80\%} possibility of going to state 5, \emph{98\%-80\%}
possibility of going to state 6. When state 5 is reached, guard
conditions demand that clock \emph{c\_c2} must be no greater than
\emph{150} (representing the system delay in 100 microseconds) and not smaller than \emph{160}  (see also Table \ref{table:t1}).

\section{Probabilistic Spatial Property Verification} \label{sec:spatial}
Probabilities in system models can affect the spatial behavior of
systems. Depending on the specification --- as provided by Reactive
Blocks ---  we can determine areas in time
and space which a system component is likely to occupy or interact
with.

\begin{figure}[tb]
\centering
\includegraphics[width=0.7\textwidth]{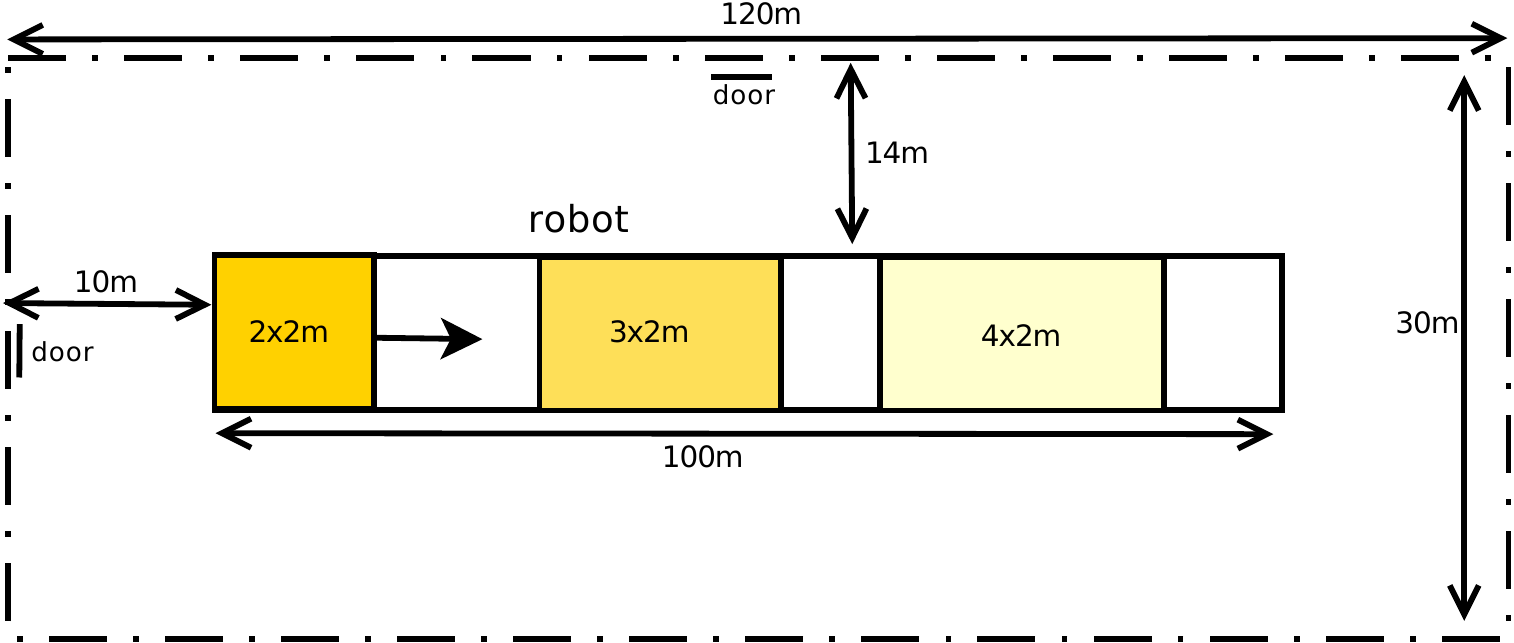}
\caption{Possible space occupation induced by unknown speed}
\label{fig:ex1ps}
\end{figure}
\begin{figure}[tb]
\centering
\includegraphics[width=0.7\textwidth]{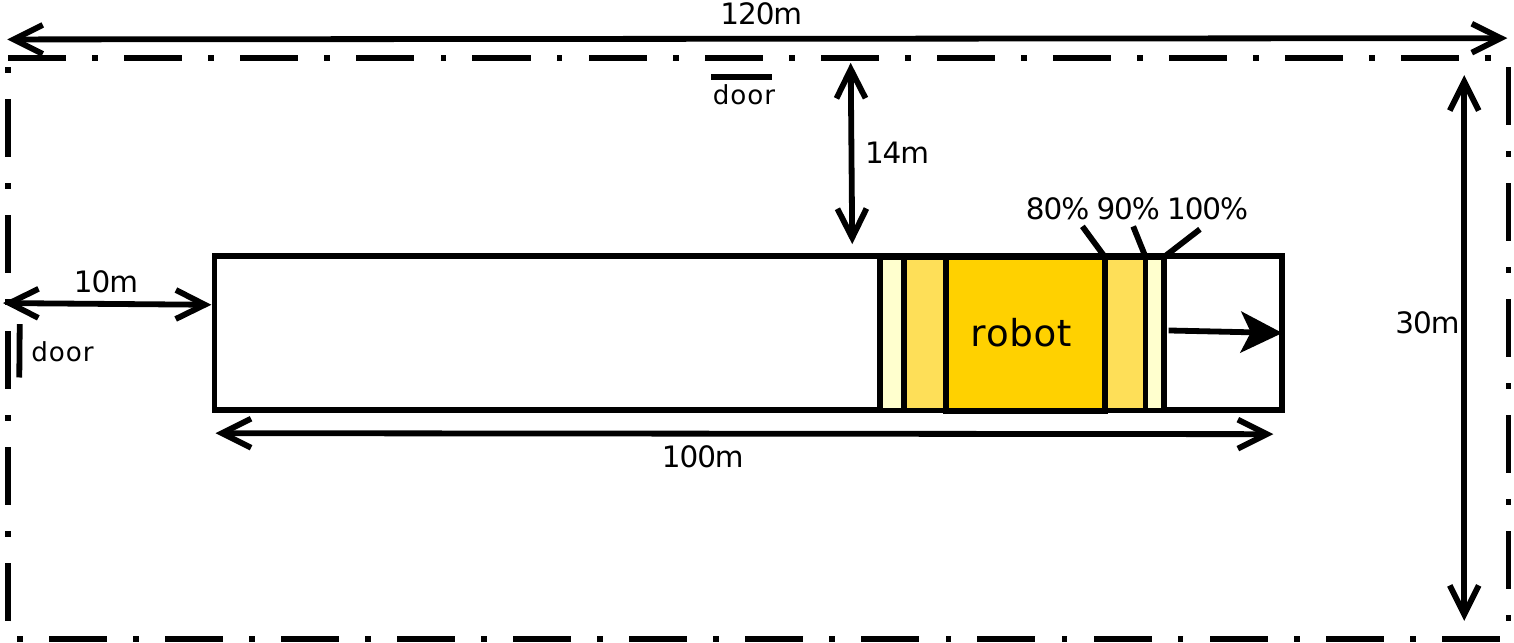}
\caption{Space occupation and probability}
\label{fig:ex1pa}
\end{figure}

In our robot system the distribution of latencies for reacting
to the detection of a human can result in different areas indicating
the possible positions of a robot for given time points, each one
associated with a probability.
Another example is varying speed. If the speed sensors of the robot are not accurate, they may come with a distribution of a possible error. Therefore the robot may accelerate to a speed slightly higher or lower than the specified 10 m/s.
As depicted in Fig.~\ref{fig:ex1ps}, this inaccuracy leads to a wider area, the robot may be in at a certain point of time, and the sizes of the areas increase over the distance the robot moves.
Furthermore, one can relate the area sizes also with probabilities.
Following the discrete probability distribution in Table~\ref{table:t1}, in Fig.~\ref{fig:ex1pa} we show the varying areas covered by the robot.
With a probability of $80~\%$ it will be within the orange rectangle at a selected point of time, with $90~\%$ in the dark yellow one and with certainty in the light yellow one.

For collision analysis we may neglect probabilities that are below a certain threshold defining residual risks that one is willing to bear.
In our example, we can prove that there is indeed a situation that a person running into the factory hall with a speed of $10\frac{m}{s}$, may hit the robot before it completely stopped.
According to the distribution in Table~\ref{table:t1}, the risk for this, however, is not higher than $5 \cdot 10^{-14}$.
Since we found out by simulating the situation that the speed of the robot at such an impact is $0.625\frac{m}{s}$ at most, the collision risk is extremely low and the impact essentially not different from the human running into a stationary object.

We implemented BeSpaceD, a tool~\cite{issec} for checking spatial
behavior of spatiotemporal systems as well as an input language for
such systems. The implementation is done in Scala and comprises
abstract datatypes that indicate spatial availability, interaction or
occupation in areas in a coordinate system for time intervals or
timepoints. It is possible to give parameterized specifications
describing non-deterministic systems and their spatial behavior.

For
this work, however, we restrict ourselves to the checking of
scenarios generated from simulation runs of Reactive Blocks
models.
Particularly, in a simulator run we stored every five milliseconds a tuple consisting of the current time stamp and the positions of human and robot in a format readable by BeSpaceD.
Thereafter, we use BeSpaceD to detect collisions and other spatial interactions for the various scenarios and probabilities.
In this way, we found out the situation mentioned above that a human indeed may collide with the robot before it has stopped.
We are able to learn such space-related safety issues already while modeling the system.
This makes it much easier to adapt the system functionality or to impose stricter real-time properties if non-bearable situations were detected.

The specification of each spatial entity in a scenario has the
form: \\ \\
\indent $time = t$ $\longrightarrow$ \hfill  \\
\indent\indent occupied spatial area  with probability $p$ $\wedge$ ...  $\wedge$
occupied spatial area  with probability $p'$ \\
\indent $time = t + 1$ $\longrightarrow$ \\
\indent\indent occupied spatial area  with probability $p$  $\wedge$  ...  $\wedge$ occupied spatial area  with probability
$p'$ \\
\begin{center}
....
\end{center}

\indent $time = t + n$ $\longrightarrow$ \\
\indent\indent occupied spatial area  with probability $p$  $\wedge$  ...  $\wedge$
occupied spatial area  with probability $p'$ \\ \\
Probabilities are treated as attributes to an occupied spatial area.
Different means to check spatial properties --- here collisions --- formulated
over these inputs are provided in BeSpaceD and are based on SAT and
SMT solving and direct Scala implementations.

\section{Probability Distributions as Verification Results} \label{sec:result}
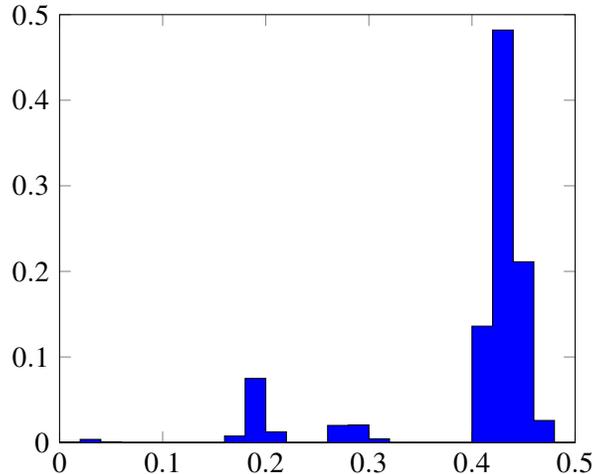
\begin{figure}[tb]
\centering
\begin{tikzpicture}
\begin{axis}[ymin=0,ymax=0.5,enlargelimits=false]
\addplot
	[const plot,fill=blue,draw=black]
coordinates
{
 (0.0000,4.0000000000000013E-4)
 (0.02000,0.0036999999997500014)
 (0.04000,8.999999977499992E-4)
 (0.06000,0.0)
 (0.08000,0.0)
 (0.100000,0.0)
 (0.120000,0.0)
 (0.140000,0.0)
 (0.160000,0.007599960000250001)
 (0.180000,0.07488979000200004)
 (0.200000,0.01251016349812517)
 (0.220000,0.0)
 (0.240000,0.0)
 (0.260000,0.020000000009000263)
 (0.280000,0.02060000672468698)
 (0.300000,0.0043899797636878235)
 (0.320000,1.0020261812682119E-5)
 (0.340000,0.0)
 (0.360000,0.0)
 (0.380000,0.0)
 (0.400000,0.1359999975002503)
 (0.420000,0.48199882466624966)
 (0.440000,0.21120057744143772)
 (0.460000,0.02578809163387452)
 (0.480000,0.0)
 (0.500000,0.0)
 }
	\closedcycle;
\end{axis}
\end{tikzpicture}
\caption{Probability density function for the system}
\label{fig:system_probability}
\end{figure}

We verify probabilistic properties based on the probabilistic timed
temporal logic PTCTL \cite{Par13}. The probability operator $P_{=?}$
allows reasoning about numerical probabilistic values, and is supported by the so called stochastic games engine \cite{KNP09d} in the verification tool PRISM.
The results of our property verification are probability
distributions. Figure~\ref{fig:system_probability} shows the
non-accumulative probability distribution of the overall system
reaction time displayed in a histogram. The \emph{x} axis represents
time values with a frequency of 0.02 s. The \emph{y} axis shows the
numerical probability value.

In our
analysis, we verified
a set of probabilities expressed in PTCTL as follows:
\begin{displaymath}
P_{=?}[F_{{\leq}T} ~ ''target'']  (T \in [0.0,...0.5])
\end{displaymath}
In the above temporal specification formula, the operator $F$ is a path
operator that equals to \emph{eventually} in LTL and can be used
inside the $P$ operator.  The pattern $F_{{\leq}T}$ stands for  ``\emph{within T time unit}''. The logical expression \emph{``target''} inside the PTA models that the parallel composed real-time probabilistic actions are indeed executed.
The formula expresses the possibility that within T time units, the labeled actions are achieved.

Important for us are questions like: Is  the robot
reaction time no more than a designed safe time limit of 0.46
s? Checking that a reaction time of 0.46 s is enough to avoid a
collision is discovered by simulation using Reactive Blocks and
BeSpaceD. With PTCTL we
check this property using the following formula:
\begin{displaymath}
P_{=?}[F_{{\leq}4600} ~ ''target'']
\end{displaymath}
The result indicates that when a human enters the monitored area, the robot reacts within $0.46~s$ with the probability of 99.99874114988752~\%.
Due to the potential severity of a collision, this number does not seem sufficient.
We discussed in Sect.~\ref{sec:spatial} that the impact will not be serious with a reaction time of $0.5~s$.
Of course, if the robot controller reacts within $0.47~s$, the maximum speed at impact will be even lower (i.e., $0.125\frac{m}{s}$ at most according to our simulations) and the risk of injuries is seen to be remote.
Considering this fact, we assume that the layout of our example system as sufficiently safe.

\section{Related Work}\label{label:relatedwork}

Here, we present related work regarding formal methods, safety
analysis, probabilities and spatial verification.

In \cite{HJ94} an algorithm is presented to check whether a given Markov Chain
satisfies formulas of Probabilistic Computation Tree Logic (PCTL).
PCTL and related verification techniques are typically
applied to analyze the reliability of timed systems. The work
presented in \cite{KNSS02} presents an algorithm for the
verification of probabilistic real-time systems annotated with
discrete probability distributions. This can be seen as a starting
point for our work.

Safety analysis is a critical phase in the development of the systems
we are aiming at. In \cite{dema08}, a solution for incorporating
safety requirements in software architecture is provided. Means facilitating
Model-Driven Architecture based development and safety analysis are presented.
The work presented in \cite{Gr05} aims at providing a modeling
framework for the hazard analysis of component-based systems. The
authors  intended to include traditional hazard analysis techniques,
e.g., Fault Tree Analysis (FTA), but found its inappropriateness for
component based systems. Thus, new techniques like State Event Fault
Tree (SEFT) are proposed. The new proposed model is suitable for
describing stochastic behaviors with the help of existing tools such as
Matlab/Simulink. These can be extended to support such models.
In \cite{EPTCS117} an extension of concurrent Kleene algebras is
provided. An axiomatisation for probabilistic, concurrent and
nondeterministic systems is presented, and the simulation and verification of probabilistic automata can be carried out.
In the past we have studied the application of probabilistic models in
a theorem prover for guaranteeing fault-tolerance of embedded systems
\cite{opodis}.

A process algebra-like formalism for describing and reasoning about
spatial behavior has been introduced in \cite{cardelli03,cardelli04}. Process algebras come with a precise formal
semantics  and target the specification of highly
parallel systems. Another logic-based approach to describe spatial areas is the Region Connection Calculus (RCC) \cite{Bennett}. It includes
spatial predicates to describe the separation and connection of
areas. The area of hybrid systems has features the development of different
tools for reasoning and verification. SpaceEx \cite{frehse2011} allows
the modeling of continuos hybrid systems based on hybrid automata. It
can be used for computing overapproximations of the space occupied by
an object moving in time and space.
Additionally, it is possible to model spatial behavior in more general
purpose oriented verification tools in Hybrid systems (e.g., \cite{keymaera}).

\section{Conclusion}\label{label:conclusions}

In this paper, we proposed an approach for the model-driven development of
probabilistic real-time systems with highly reusable compositional
building blocks that incorporate discrete probability distributions for
describing stochastic time behaviors. These extensions are used to
predict system performance as well as probabilistic safety properties. In addition, we  established a development tool set both for temporal and spatial probabilistic behaviors of systems.
The Probabilistic Real-Time External State Machines (PRTESM) of building blocks give a straightforward
view of stochastic real-time behaviors of component-based
systems. Software architects and safety engineers can use this for verification and analysis.
The limitations of the approach is that some intral or inter components' communications are, instead, described as messages passed and received in UML, such that such behaviors can not be captured by ESMs. Also, whether this approach is applicable  depends on the abstraction level.
In the future, we want to further study and emphasize the compositionality of probabilistic spatial behavior definitions, i.e., of systems composed of appropriately logically detailed subsystems. We plan to advance the description logic as well as the introduction of a probabilistic spatial behavioral type infrastructure that extends previous work \cite{behtosgi,isolavision12}.

{\bf Acknowledgments} We like to express our thanks to Song Zheng Song from the National University of Singapore (NUS) for the useful discussion during the research work.

\bibliographystyle{eptcs}
\bibliography{reference}
\providecommand{\urlalt}[2]{\href{#1}{#2}}
\providecommand{\doi}[1]{doi:\urlalt{http://dx.doi.org/#1}{#1}}

\end{document}